\documentclass[conference,letterpaper]{IEEEtran}
\IEEEoverridecommandlockouts

\usepackage{cite}
\usepackage{amsmath,amssymb,amsfonts}
\usepackage{algorithm}
\usepackage{algpseudocode}
\usepackage{textcomp}
\usepackage{xcolor}
\usepackage{url}
\usepackage{adjustbox}
\usepackage{multirow}
\usepackage{subfigure}
\usepackage{enumitem}

\newif\iffinal
\finaltrue

\iffinal
  \newcommand\raj[1]{}
  \newcommand\joaquin[1]{}
  \newcommand\ganesh[1]{}
  \newcommand\todo[1]{}
\else
  \newcommand\joaquin[1]{{\color{blue}[Joaquin: #1]}}
  \newcommand\raj[1]{{\color{olive}[Raj: #1]}}    \newcommand\ganesh[1]{{\color{cyan}[Ganesh: #1]}}
  \newcommand\todo[1]{{\color{red}[TO-DO: #1]}}
\fi

\def\BibTeX{{\rm B\kern-.05em{\sc i\kern-.025em b}\kern-.08em
    T\kern-.1667em\lower.7ex\hbox{E}\kern-.125emX}}
    
\begin{document}

\title{La R\'esistance: Harnessing Heterogeneous Resources for Adaptive Resiliency in 6G Networks}

\author{\IEEEauthorblockN{Ganesh C. Sankaran}
\IEEEauthorblockA{\textit{Information Sciences Institute} \\
Los Angeles, CA, USA \\
gsankara@isi.edu}
\and
\IEEEauthorblockN{Joaquin Chung}
\IEEEauthorblockA{\textit{Argonne National Laboratory} \\
Lemont, IL, USA \\
chungmiranda@anl.gov}
\and
\IEEEauthorblockN{Raj Kettimuthu}
\IEEEauthorblockA{\textit{Argonne National Laboratory} \\
Lemont, IL, USA \\
kettimut@anl.gov}
}

\maketitle

\begin{abstract}
Recent years have seen more critical applications designed to protect human lives (e.g., environmental sensing, emergency response, and tactical defense) being deployed over wireless networks. These critical deployments expect higher data rates, ultra-low latency, and ultra-high reliability. 6G wireless networks are expected to fill the gap in terms of the first two aspects (i.e., higher data rates and ultra-low latency), however providing ultra-high reliability is a wide open challenge. All is well when everything works but when there is a failure or a security attack, the entire system collapses exposing the associated human lives to imminent danger. Avoiding this requires the strongest of assurances that safety and security aspects are protected no matter what. Large scale critical applications are waiting for this piece of the puzzle to be solved.
At this juncture, we envision the bold theme of La R\'esistance 6G (LR6G) that would pave the way for deploying mission-critical applications and services over 6G networks. It aims to achieve ultra-high reliability and resiliency to any disruptions, be it failures or security attacks. A few disruptions are easy to handle such as a cloud VM or primary link failure. In both of these cases, applications can be restored by activating the standby resource. However, some disruptions can be detrimental such as when a cut-vertex fails or when the disruption leaves the critical application to fail without access to a standby resource.
These critical applications (e.g. Smart Manufacturing, Smart City, Ocean monitoring, Wildfire monitoring, etc.) are highly distributed in nature. They must continue to deliver their mission objectives during a disruption to protect human lives. In this paper, we present our LR6G vision and outline the challenges towards achieving this vision.

\end{abstract}



\section{Introduction} \label{sec:introduction}

The sixth generation (6G) of wireless mobile networks are expected to have key performance indicators such as 1~Gbps data rates (per user), ultra-low latency (1 ms or less), massive numbers of devices, and ultra-high reliability (99.99999\%)~\cite{6G-vision}.
6G will also offer significant improvement in programmability and quality of service (QoS) by leveraging technologies such as software-defined networking (SDN), network function virtualization (NFV), multi-access edge computing (MEC), in-network computing, dynamic orchestration, and machine-learning/artificial-intelligence (ML/AI).
All these features make 6G networks a complex infrastructure that provides a plenty of opportunity to incorporate non-traditional resiliency capabilities in the network for networked applications.

Systems built to protect human lives are increasingly deployed over communication networks. 
Examples of these systems are the smart city, smart agriculture, power grid, autonomous vehicle traffic systems, tactical defense networks, and emergency response systems.
When communication breaks or is delayed, it puts human lives in danger. 
For example, delay in notifying a frequency drift in a smart grid network will not start the power generation at the right time and cause the entire grid to collapse in a ripple effect. 

The current generation of networks (are expected to) meet their mission objectives under normal operating conditions, but we envision 6G-enabled systems (with an architecture that exploits the advanced features intelligently such as the one proposed here) will be capable of meeting mission objectives even during disruptions.
In order to handle a wide variety of adversarial and failure conditions, resiliency should be a pillar of 6G system design and architecture.
In the context of 6G systems, resiliency can be defined as the ability to survive, gracefully adapt to, and rapidly recover from malicious attacks, component failures, and natural and human-induced disruptions.

To provide resiliency, the traditional approach is to replicate a component so that when one instance goes down, the rest can handle new requests~\cite{9279239}.
However, replication is limited to homogeneous stack environments containing similar hardware, operating system, libraries, etc. 
This work takes a step beyond the current state-of-the-art to restore the critical functionality. We propose to restore critical functionality by adapting it on to the available non-homogeneous resources in the device-to-edge-to-cloud continuum of 6G networks. \raj{edge-to-cloud will not be 6G?}

\begin{figure*}[htb!]
    \centering
    \subfigure[Discrete functions mapped onto discrete resources]{
    \includegraphics[width=0.48\textwidth,trim=50 0 200 80, clip]{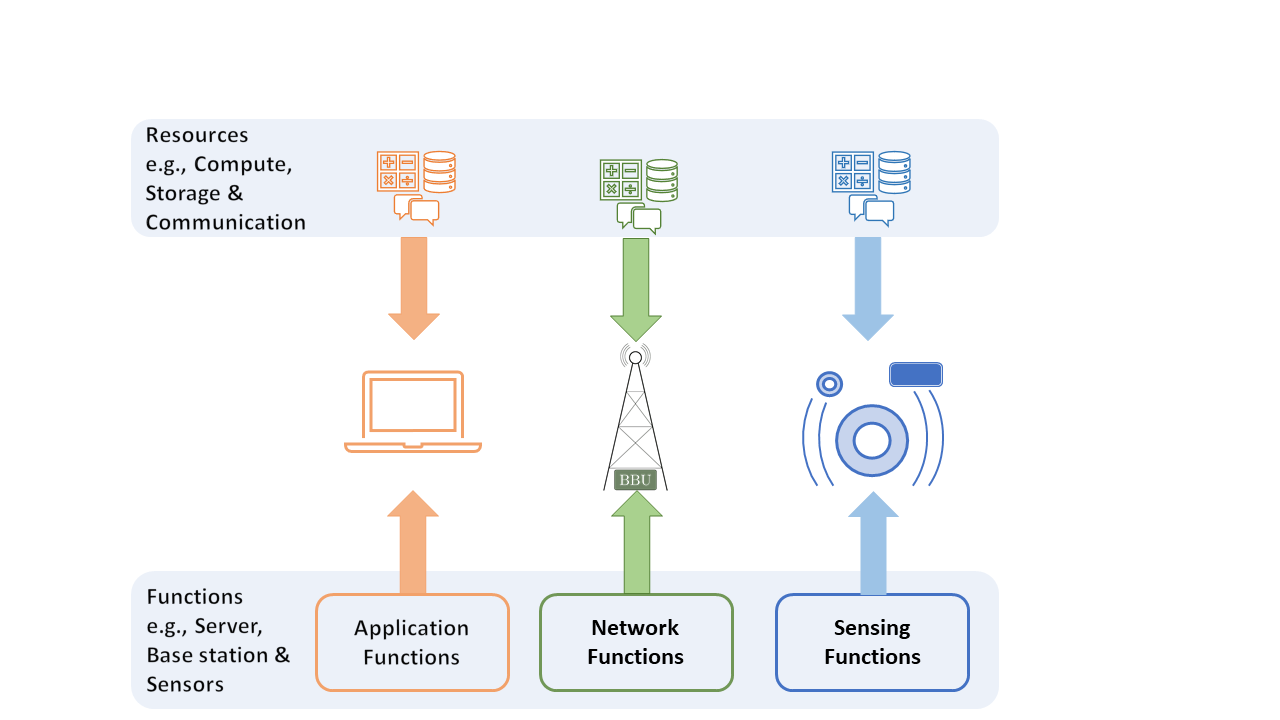}}
    \subfigure[Unified functions mapped onto unified resources]{
    \includegraphics[width=0.50\textwidth,trim=50 0 200 100, clip]{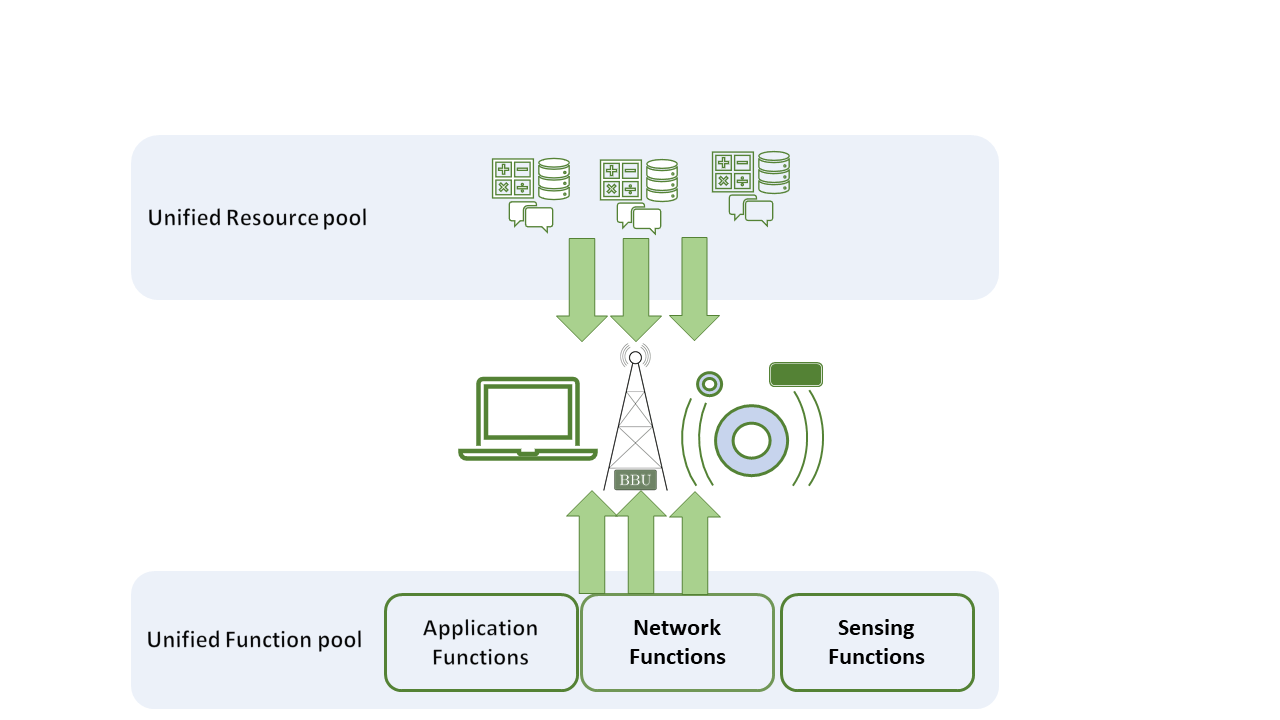}}
    \caption{Unification vision.}
    \label{fig:hmap}
\end{figure*}

We envision a La R\'esistance 6G (LR6G) network with adaptive resiliency. To motivate this, let us consider the following example. A critical function/service hosted on the C-RAN is disrupted due to a security (say reflective DDoS) attack. The Edge-to-Cloud network is flooded and is dropping all benign packets as suspicious. The only resilient option is to harness a few small IoT devices to put up a valiant fight against the adversary.

5G network relaxes the association of network function to a part of the network (e.g., 5G Core or NG-RAN). 5G operator chooses a placement that optimizes their objective function. The 5G network does not yet understand the notional loss resulting out of the service disruption. 5G network handles the following disruptions in this example scenario:
\begin{enumerate}[label=\alph*)]
    \item Connectivity disruption case: The 5G network or the edge network attempts to establish alternate  connectivity. This is part of path diversity and multi-homing approaches.
    \item Cloud resource disruption case: Cloud manager migrates the function to a new cloud resource. This is more a disruption-centric best effort approach. When notional loss is high, restoring the service becomes super-critical. For instance, service disruption in a nuclear power plant would cause a radiation exposure threat in the whole region e.g., Fukushima disaster. With 6G, IoT devices are expected to take up more and more mission-critical that tasks directly impact human lives.
\end{enumerate}

LR6G network is aware of its criticality and adapts itself to restore services in non-traditional ways. First of all, LR6G network expands its scope beyond network functions and is more system- and service-centric. LR6G has the following distinctive characteristics:
\begin{enumerate}
    \item It creates a unified resource pool of the available resources be it network equipment, storage equipment, or IoT devices without loosing visibility into their individual  capacity, strengths and weaknesses.
    \item It creates a unified programmable platform out of heterogeneous resources to meet desired performance and security objectives.
    \item It has the ability to slice and dice the functionality to fit individual resource capacity constraints. While doing so results in an increase in the attack surface. The unified platform will have the ability to create suitable checks and balances to identify future issues arising due to any weak links across the system.
\end{enumerate}

The rest of the paper is organized as follows: Section~\ref{sec:background} provides background on resiliency strategies for distributed functions. Section~\ref{sec:prop-work} presents our proposed La R\'esistance 6G (LR6G) architecture.
Finally, Section~\ref{sec:conclusion} presents our conclusions.
\section{Background} \label{sec:background}

Self-healing networks heal themselves without a human intervention \cite{wang2021sdn}. As and when disruption occurs, the network detects the situation and makes the necessary changes in an automated manner to heal itself.  Drawing a cue from self-healing networks, our LR6G system will restore its functionality without human intervention.

Today, networked applications (such as smart city, smart agriculture, autonomous cars) are built to meet specific societal objectives with components distributed across the network~\cite{9173706,kirimtat2020future,friha2021internet}. 
A set of distributed functions that is essential to meeting these objectives are henceforth known as critical functions. These functions are not only restricted to the application layer but also include security, safety, and critical network layer functions.
During a network failure, a portion of the network nodes lose communication from the rest.
Such failure may be the effect of a variety of causes including but not limited to (i) equipment misconfigurations~\cite{246314}, (ii) component aging~\cite{paing2020analysis}, or (iii) a systematic attack~\cite{iot-ddos}.
This is likely to happen more often in the far edge of wireless networks. As 6G is expected to operate in higher bands of the spectrum, signals are more directional, more susceptible to interference, requiring more intelligent management of the spectrum.

Currently, a given component (function) is mapped to its stack (underlying hardware, operating system, and software libraries) at design time and any major change is a  time-consuming effort ~\cite{chirigati2021porting} \raj{what is not altered? the function? is this true? what happens when the library or OS has to be upgraded for security vulnerability reasons and that breaks backward compatibility? Also, it is not clear what ``lifetime of the solution'' mean}. 
In the device-to-edge-to-cloud continuum of 6G, it may be difficult to find homogeneous resources. We envision to restore functionality using non-homogeneous stack. 
However, this is a challenging task as this new pool of resources may be resources constrained or may have non-traditional architecture and programming paradigms.
The following subsections provide details on our approach.

\subsection{Gradual Service Degradation} \label{sec:grad-degrad} 
\begin{figure}[htb!]
    \centering
    \vspace{0.5em}
    \includegraphics[width=0.9\columnwidth,trim=350 50 150 20, clip]{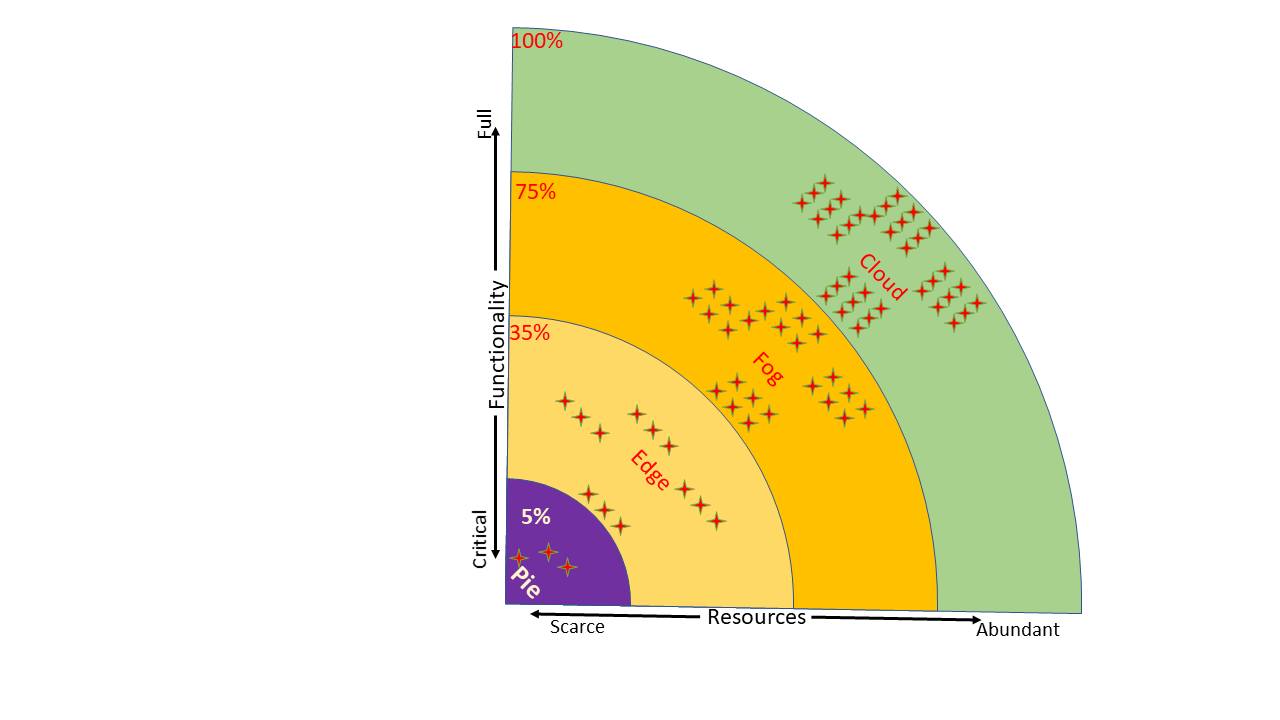}
    \caption{Gradual service degradation as a function of resource availability and functionality}
    \label{fig:tactical}
\end{figure}

Fig.~\ref{fig:tactical} represents gradual service degradation, a paradigm that networked applications could leverage. When there is cloud connectivity, sufficient resources are available to support full functionality. When connectivity to cloud fails, reachability up to Fog is present. In this situation, less than sufficient resources are available. With these resources the network architecture should lose a part of the functionality but should not cripple in terms of operation. Similarly, when fog or edge connectivity is lost, the functionality and resources are further reduced. With the available resources, the system shrinks to keep as much functionality alive. This includes the critical functionality. Based on resource availability, more high priority functionality is supported.

To take advantage of gradual degradation during a disruption, a 6G-enabled system must monitor its components for failures and security compromises, compute a sequence of functions that will meet the mission objectives, and adapt these functions to the available resources.
However, key challenges must be addressed for deploying resilient 6G-enabled systems:
\begin{itemize}
    \item State-of-the-art monitoring mechanisms can be either complex or simple, with complex mechanisms requiring considerable amount of resources and simple mechanisms lacking in accuracy. A light-weight highly accurate monitoring mechanism is required to detect performance and security anomalies in the system on a continuous basis. \joaquin{This is because of the wireless media and resource-constrained devices, we need to be more emphatic about it.} \ganesh{Sure. A point to note is the attack surface increases when distributing a function. We must emphasize that we need to solve the on-the-fly version of the problem to achieve this.}
    \item Current resiliency mechanism rely on homogeneous full-stack resources for resiliency by replication. \ganesh{Do we need to also mention Network connectivity resiliency solution space?} \joaquin{I think so, 5G already supports device-to-device communication in case of losing connectivity to gNodeB} However, this limits the number of available resources that are able to maintain critical functions operations. The Cloud has abundant homogeneous resources, but the rest of the device-to-edge-to-cloud continuum lacks abundant homogeneous resources.
    \item By making heterogeneous resources available, we could increase the pool of resources that critical functions could use to remain operative. However, converting a cloud-function to a edge-function or a in-network function is not straightforward. Furthermore, when one-to-one function-to-resource mapping is not found, the function must be collectively realized by many resources. 
    \item The problem of choosing where and how to redeploy a critical function for resiliency can rapidly increase with the number of requirements. These requirements may include meeting function resource demands with available resources; resolving any critical, security and safety function sequencing conflicts; and meeting the performance and security objectives.
\end{itemize}

\ganesh{convert this into unified resource pool}
In general, a programmable data plane pipeline may perform a function on the data stream that would have otherwise been executed on an end node.
It computes a result (e.g., future hardware mapping and placement of a function 
) that will be consumed by an application (e.g., computer vision or augmented reality).
In previous work~\cite{inc-netsoft2021} we demonstrated how to leverage in-network computing for performing scientific operations (that are not supported in network devices) using approximations in a streaming fashion.

As we will discuss in Section~\ref{sec:prop-work}, programmable data plane pipelines could be used to host critical functions in response to a disruptive event (e.g., attack or failure).
However, many challenges remain open for this technology to be a real enabler of resiliency in 6G networks and systems.
\section{Envisioned LR6G system} \label{sec:prop-work}

\begin{figure*}[htb]
    \centering
    \vspace{5mm}
    \includegraphics[width=.8\textwidth]{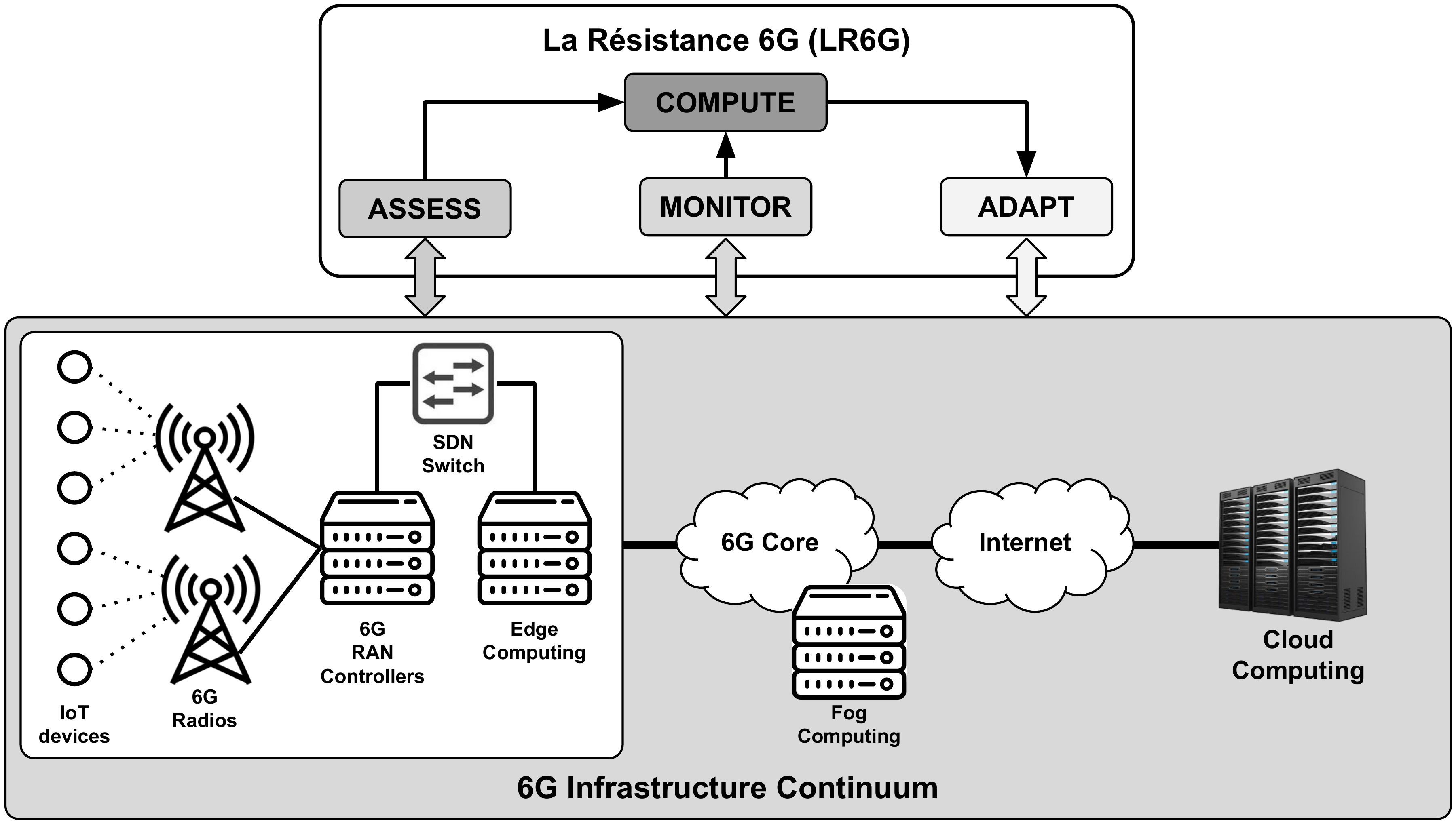}
    \caption{La R\'esistance 6G Architecture} \ganesh{should we move this to the front page to set context on disruption and recovery?} \joaquin{Perhaps, but I'm not sure bc it shows the 4 components of the architecture}
    \label{fig:arch}
    \vspace{1em}
\end{figure*}

We propose La R\'esistance 6G (LR6G), a dynamic architecture for 6G systems 
that adapts 6G networked applications to meet critical security and safety requirements during a disruption. 
We envision
\begin{enumerate}
    \item a self-restoring system that restores and retains its critical functionality in the event of a disruption,
    \item a unified pool of resources that can be mobilized in the event of a disruption,
    \item a software-defined distributed system that provides a unified programming model to program a wide variety of devices, and
    \item the ability to survive disasters (or) in other words survival to $N-\epsilon$ failure
\end{enumerate}

Being adaptive is a control loop where the current situation is sensed at first and then the system reacts to the current situation to meet the mission objectives. We partition our adaptive vision into four broad areas viz. (1) ASSESS, (2) MONITOR, (3) COMPUTE, and (4) ADAPT. Here, the first two map to static and dynamic analysis of the situation, while the last two map the system's reaction towards service restoration.

At a high level, the individual areas will encompass the following:
\begin{enumerate}[label=(\roman*)]
    \item ASSESS: assess available hardware resources (network switches, accelerators, and IoT devices) on a range of security and performance parameters. This static analysis will help in determining the capacity of the available unified resource pool.
    \item MONITOR: monitor resource utilization to detect security and performance anomalies in a granular fashion.
    \item COMPUTE: compute the optimal function sequence based on the current network state (failure and attack state), available resource capacity, the resource constraints (in terms of performance and security) and semantic relationships. This will compute the mapping of code fragments onto the available resource pool.
    \item ADAPT: We envision a unified \ganesh{or should it be common} programming model to distribute and program the critical functions onto the available resources. 
\end{enumerate}

We now present how each of these areas and associated challenges.

\subsection{ASSESS} \label{sec:ASSESS}
We envision ASSESS to provide greater visibility into the available resource pool creating a unified view. This unified view must encompass various dimensions of the resources such as power, performance and security. Understanding these dimensions will enable us  create a minimally vulnerable or minimally susceptible system.

Device layers such as hardware, operating environment, and applications may have different vulnerabilities or be susceptible to different types of failures. The key challenge is to have an accurate view of these attributes for every device. This is combinatorial and it is difficult to assess every device or every (hardware, OS, application) combination. An orthogonal view of hardware, OS, and application is useful to represent a device without an individual assessment but such a view will not be able to capture cross-layer vulnerabilities. For instance, studying the hardware (say ARM core) in isolation will be able to cover a variety of devices using this hardware model. Similarly, study of OS (say tinyOS) will cover a wide range of devices running this OS. This study is insufficient to cover the vulnerabilities with the ARM-tinyOS combination. With the volume and heterogeneity of IoT devices, the scale of this challenge balloons up.

\begin{figure}
    \centering
    \includegraphics[trim=75 40 420 10,clip,width=0.8\columnwidth]{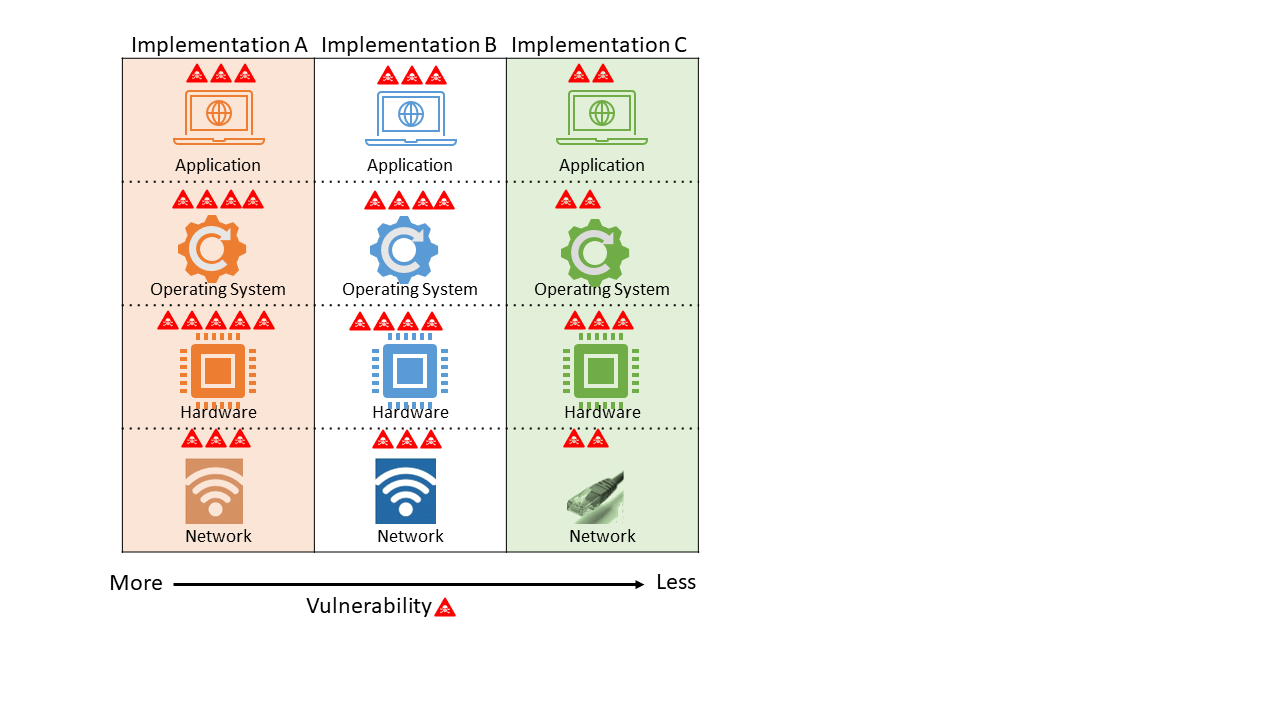}
    \caption{Quest for the minimalist vulnerable implementation.} 
    \label{fig:vul-layer}
\end{figure}

Fig.~\ref{fig:vul-layer} presents an example in which three system implementations (shown in different colors) are considered. The system on the left is built on top of hardware $h_a$, operating system $o_a$, and in programming language $l_a$. Each of these layers is known to have some vulnerabilities. A higher number of red triangular skull icons in the layer means that more vulnerabilities are known. For instance, hardware $h_a$ is known to have more vulnerabilities than hardware $h_c$. Thus, hardware $h_c$ is considered more secure. Both performance and security must be studied to see how hosting specific functions or a part of it can improve or degrade the critical application.

New vulnerabilities are reported every day. ASSESS should keep pace with new vulnerabilities and in the best case should be able to identify deployment specific vulnerabilities. 
Furthermore, the wireless channel of 6G may be susceptible to new type of vulnerabilities in the physical layer.

Another key challenge \ganesh{may be we should list the challenges upfront} \joaquin{agree} is isolation. The chosen hardware that is adapted will host critical functions in addition to the existing functions it is already hosting. Any problem encountered by the new function must not impact existing functions it is hosting and vice-versa. Isolation helps in containing any further damage to the available resources. Most devices other than cloud computing devices do not support isolation or virualization. Moreover IoT devices are resource-constrained and it is difficult to realize the desired level of isolation over them.

As already mentioned, our vision is to restore the critical functions. This involves mobilizing IoT devices and/or other network resources such as RAN that are otherwise not meant to run a critical function. A few vendors lock the programmability for security reasons. The programming model of these devices are very different.



Once we detect a vulnerability in a device or a system, how quickly can we plug it is a key challenge. In this context, event based programming models used in DataOps pipelines are very agile. The entire pipeline can be modified and upgraded on the go. A similar capability is envisioned.

\subsection{MONITOR} \label{sec:MONITOR}
This area keeps the system grounded to the current network state.
Without sufficient monitoring, a disruption of critical functionality on a 6G system may go undetected.
Response time largely depends on the time taken by monitoring considering rest of the times to be a constant. The aim is to detect the system disruption in real-time or much better if it is predicted. This is a huge challenge and a responsibility on monitoring area.

Detailed monitoring is achieved at the increase in cost of adding intrusive instructions to the devices. Many anti-virus software slow down compute hardware by introducing more monitoring instructions per executed instruction on an average. As we consider resource-constrained IoT devices, monitoring must be timely but at the same time resource friendly. This is a challenge.

Accuracy of detection must be high. Otherwise, monitoring loses its credibility. False positives are seen as a source of DoS attack on the resources. The same level of accuracy must be maintained even when prediction mechanisms are used. False negatives both impact the system detection ability and will leave the system in a broken state.

When the system is adapted to restore critical functions, new monitoring hooks will be required. For instance, a critical application when hosted in a cloud environment can afford sophisticated monitoring ability. However, when the same functionality is mapped to a resource-constrained device such as an IoT device, the same monitoring hooks will not just fit. It is challenging to reintroduce equivalent resource friendly monitoring without manual intervention.

Let us consider a situation where monitoring is mapped onto device 'X' that is along the network path. However, when the system is adapted, where functionality is hosted on a different device, this device 'X' may not be on the network path. Thus a realignment is required as and when the functionality mapping is revisited. All monitoring functions and their mapped resources must be well-known. It must be budgeted for whenever the functional mapping is computed. Realignment is challenging without manual intervention.

\joaquin{Can we add something about network usage? Specially if the interface is wireless, can we add something about the status of the wireless channel?} \ganesh{Yes}

\noindent\textbf{Detect system level anomalies:} As noted already, it is challenging to define mechanisms that detect anomalies at a granular level without incurring high cost. Monitoring response time to detect anomalies \cite{9006046,8116438} is at one end of the spectrum. Producing a digest capturing every instruction executed along with CPU cycles is at the other end of the spectrum. Response time is a coarse measure that enables identification of huge deviations and the digest is the most granular that can help identify even minute deviations. However, the cost of instruction level digest is high. It is a challenge to choose the right mechanism with the right balance. 

\noindent\textbf{Detect disruptions within a distributed function sequence:} When the resource demands do not fit a single node, the function is distributed across nodes (\joaquin{see ADAPT for details?)}. When realizing a function across nodes, monitoring must be strengthened to identify failures and potential disruptions at the earliest and to restore the system. 
These disruptions may include security vulnerabilities or the 6G wireless channel.
Designing mechanism to obtain information on forwarding decision for packet flows is a promising approach to build a quick isolation mechanism that is capable of identifying the faulty component in a sequence of distributed functions~\cite{sankaran2010obtaining}.
Furthermore, keeping track of the wireless network status and mapping it to the services impacted would be the key for our vision.

\subsection{COMPUTE} \label{sec:COMPUTE}
This area takes all available input factors to compute the mapping of critical functionality to available resources. This computation is trivial when the given resource pool is homogeneous and individual resource capacity is at least the maximum resources required to host any critical functionality. 

Here are the key challenges that must be considered during this computation.
\begin{enumerate}
    \item As we discussed already, a LR6G resources exhibit more heterogeneity. This obviously indicates that their resource capacities are not equal. The individual resource capacity and resource requirements of the critical functionality have no relation. In many cases, functional requirements are more than capacity. This problem is no longer trivial and must consider fragmentation.
    \item The side-effect of fragmentation is distribution. Attack surface increases when the functionality is distributed. Attack surface increases when a function is distributed. The successor device should be able to identify whether the information received is really from its predecessor device and it is valid input. Predecessor device can get corrupted or spoofed. So, additional monitoring hooks to detect performance and security compromises such as execution time, proof of benign computation to detect any malicious code traversal, and proof of benign data in terms of validation of its input are required. 
    \item Heterogeneity increases the dimensions of the problem manifold. Volumetric attributes such as capacity constraint make this problem look more like knapsack. Categorical attributes such as resistance to security attacks makes it more like a Max-SAT problem. The state in terms of the security attacks that the system is facing is critical. One manifestation of the problem is to associate individual SAT clauses with different weights and maximize the satisfied weights.
    \item Critical  safety and security functions are mapped to fractional variables that add up to one when a function is fully implemented. The corresponding constraints must be satisfied outside the SAT clauses for security. Performance metrics must be satisfied as well. Metrics such as latency are non-linear in a multi-commodity flow formulation.
    \item Multi-hop wireless topology must be considered when distributing functions across a set of devices from a proximity point of view. Renewable resources do not impact system lifetime. Similarly, on-tap resources if any available must be factored into the computation.
\end{enumerate}

A formulation considering the above challenges will provide insights into the time complexity of the problem. We conjecture that this computation problem is NP-Hard. Response time includes computation and computing a feasible solution over a longer duration in the order of minutes can worsen the situation. Just solving the security part of the problem involving a large number of categorical variables will take much more time. Suitable heuristics must be defined to find a feasible solution quickly.

\subsection{ADAPT} \label{sec:ADAPT} \ganesh{TODO} 

This area starts where COMPUTE ends. It articulates challenges involved in taking the functional mapping result from COMPUTE to create the binaries and configuration all the way up to complete the deployment. 
\begin{enumerate}
    \item The new function that is added to the devices must not impact their existing functionality in terms of performance or security or availability. Many resource-constrained devices are used and it is not feasible to build a strong isolated environment such as a sandbox or a container or a virtual machine that does not impact the rest of the device.
    \item Heterogeneity has been mentioned as a challenge in other areas but here the lack of uniform programming and deployment model creates a host of issues. Creating a binary for a device without human intervention becomes a huge challenge. The code written for a non-real-time OS does not go well with a real-time OS target. We lack a common compiler toolchain that can create binaries for all device targets. For many small devices, OS and application are built together into a monolithic binary. Even standard library functions (e.g., standard I/O) are not available across all targets. Some platform compiler toolchains are not available publicly. In a few cases, the programming model is different for the hardware platform. Specifically, hardware that are based on GPU, FPGA, and NPU based devices. Usually portable libraries are used to be able to port the code to similar target devices but porting them across all targets found in a 6G environment is an open challenge.
    \item Configuration interfaces are not uniform. The configuration interface for an IoT device is not similar to the BBU configuration interface. Especially given the volume and wide variety of IoT devices, lack of common configuration interfaces is still an unsolved challenge.
\end{enumerate}

\section{Conclusion} \label{sec:conclusion}
We motivated the need for critical systems that impact human lives more than any natural disaster such as life-critical systems used in hospitals. These systems must survive no matter what.
We then presented our vision of La R\'esistance 6G (LR6G) system viz. the vision for self restoration of critical services, ultra-high reliability, unified view of available resources, uniform programming model and deployment model.
Finally, we presented challenges in realizing this vision. Some challenges are wide open and involve multiple stakeholders. The entire 6G ecosystem must come together to solve these challenges.

\section*{Acknowledgments}

\bibliographystyle{IEEEtran}
\bibliography{bibliography.bib}

\end{document}